# Nonlinear Hall effect as a signature of electronic phase separation in the semimetallic ferromagnet EuB$_6$

*Xiaohang Zhang*[1†], *Liuqi Yu*[1], *Stephan von Molnár*[1], *Zachary Fisk*[2], and *Peng Xiong*[1]

[1] Department of Physics & MARTECH, Florida State University, Tallahassee, Florida 32306, USA

[2] Department of Physics, University of California, Irvine, California 92697, USA

[†]present address: Department of Physics, University of Maryland, College Park, Maryland 20742, USA

**ABSTRACT**  This work reports a study of the nonlinear Hall Effect (HE) in the semimetallic ferromagnet EuB$_6$. A distinct switch in its Hall resistivity slope is observed in the paramagnetic phase, which occurs at a *single critical magnetization* over a wide temperature range. The observation is interpreted as the point of percolation for entities of a more conducting and magnetically ordered phase in a less ordered background. With an increasing applied magnetic field, the conducting regions either increase in number or expand beyond the percolation limit, hence increasing the global conductivity and effective carrier density. An empirical two-component model expression provides excellent scaling and a quantitative fit to the HE data and may be applicable to other correlated electron systems.

The intricate balance of various competing interactions in complex materials has produced some spectacular physics such as high temperature superconductivity (HTS) in cuprates and colossal magnetoresistance (CMR) in magnetic semiconductors. A ubiquitous characteristic of such strongly correlated materials is a rich phase diagram in which many phases exhibit intrinsic electronic (nonchemical) phase separations, and consequently, some of the phase transitions are percolative in nature. Such nanoscale phase separations are thought to play critical roles in the emergence of HTS [1] and CMR [2], and they have been probed with a variety of direct and indirect experimental techniques.

The Hall effect (HE) is one of the most frequently performed electrical measurements in condensed matter physics. For a simple single-band material, the Hall resistivity as a function of the magnetic induction is a straight line whose slope yields the type and density of the charge carriers. The carrier density also corresponds to the volume enclosed by the Fermi surface and hence is largely temperature independent. The linear field dependence of the Hall resistivity breaks down, however, in a variety of circumstances such as in materials with a multi-band Fermi surface [3], anomalous Hall effect (AHE) in magnetic materials [4], and magnetic field induced phase transition [5] or charge localization [6] in semiconductors. Recently, a new type of nonlinear HE was observed in the heavy fermion metal YbRh$_2$Si$_2$ [7], where each Hall resistivity curve at different temperatures takes on two distinct slopes at low fields and high fields respectively, and the crossover field increases linearly with temperature. The authors suggested that the switch in the Hall resistivity slope corresponds to a sudden change of the volume of the Fermi surface and essentially relates to the existence of quantum criticality.

In this letter, we present an analysis of a similar switching of the Hall resistivity slope in the paramagnetic phase of the ferromagnetic semimetal EuB$_6$. Most significantly, two important features which relate the nonlinear HE to the magnetic state of the material have been identified:

1) the linear temperature dependence of the crossover field extrapolates precisely to the paramagnetic Curie temperature; 2) for a wide range of temperatures the switching occurs at a *single critical magnetization*. We interpret these features as distinct indicators of electronic phase separation and a percolative phase transition.

$EuB_6$ has a simple cubic structure and remarkably rich magnetic and transport properties. Specific heat [8-10] and electron transport [9,10] measurements revealed two consecutive phase transitions at 15.3 K and 12.6 K, respectively. In the ferromagnetic phase, an intrinsic semimetallic band structure was evidenced by quantum oscillation measurements [11]. Recent Andreev reflection and low-field HE measurements further indicate the semimetallic band structure likely consists of a completely polarized valence band and a non-polarized conduction band in the *ferromagnetic* phase, and a localized hole band in the *paramagnetic* phase [12].

Fig. 1(a) shows the overall features of the nonlinear HE measured on a $EuB_6$ single crystal (a $4.0 \times 2.2 \times 0.1$ mm$^3$ platelet) in fields up to 7.5 T for various temperatures. At temperatures above $T_C$, the Hall resistivity curves have two distinct slopes: a larger slope for low fields and a smaller slope for high fields. The inset of Fig. 1(a) shows the high-field (squares) and low-field (circles) slopes for different temperatures. Evidently, in the ferromagnetic state the Hall resistivity slope takes on the small value over the entire field range, while in the paramagnetic phase there are two distinct values for the slopes at high and low fields respectively with the high-field slopes consistent with the Hall coefficients in the ferromagnetic phase as indicated by the solid line. This nonlinear Hall resistivity is not related to the conventional AHE. This is evidenced directly from the lack of any simple correlation between the Hall resistivity and the magnetization (Fig. 1(b)); especially, the magnetization is essentially featureless at the switching fields for the Hall resistivity. Moreover, an estimate of the AHE coefficient ($R_s$) from the scaling between the zero-field Hall coefficient and resistivity ($\rho$ or $\rho^2$) [4] yield values at least an order of magnitude smaller than the normal Hall coefficient ($R_n$), consistent with results in the magnetic semiconductor EuS [13] where the spin-orbit interaction is very weak. Instead, the slope change originates from a change of $R_n$, that is, a change in the *effective* carrier density or the band structure with increasing magnetic field.

To uncover the underlining mechanism of the nonlinear HE, linear fits to the low-field and high-field Hall resistivities are separately performed for each paramagnetic temperature. Extrapolations of the two linear fits then result in a crossover field $B_C$ (inset of Fig. 2(a)), which depends linearly on temperature as shown in Fig. 2(a). Remarkably, the data extrapolate to a temperature of 15.6 K, which coincides with the paramagnetic Curie temperature, $\theta$, obtained from the magnetic susceptibility measurements, as shown in Fig. 2(b). This is an important result which suggests a constant value for $B_C/(T-\theta)$, independent of temperature. Since $B/(T-\theta)$ is proportional to magnetization $M$ for the paramagnetic phase (Curie-Weiss law), the data point to an important conclusion: *the observed switch in Hall resistivity slope occurs at a single critical magnetization, $M_C$, at all temperatures*.

To quantify the role of magnetization, we plot the Hall resistivities measured at paramagnetic temperatures as a function of $B/(T-\theta)$, *i.e.* the effective magnetization at field $B$ for the paramagnetic phase. In order to retain the same slope, the Hall resistivities are rescaled by the same factor of $1/(T-\theta)$. Remarkably, the rescaled data for a broad temperature range all collapse onto a single curve, as shown in Fig. 2(c), providing compelling evidence for its direct correlation with magnetization rather than applied magnetic field or temperature. To further verify this statement, Hall measurements have been performed in which the predominant field component is in the plane of the crystal ($B_{\parallel}$); only a small out-of-plane component is used to determine the Hall coefficient. The results are fully consistent with those from the perpendicular field measurements (Fig. 1(a)): the Hall coefficient exhibits a distinct change of similar magnitude at similar (parallel) field strengths. Scaling of the data with $B_{||}$ is shown in Fig. 3(a). The evidence for a critical magnetization is unambiguous, since the scaling is with $B_{||}$.

The excellent scaling shown in Fig. 2(c) suggests that the observed nonlinear HE can be described by only three parameters: the low-field Hall resistivity slope, high-field Hall resistivity slope, and magnetization. We consequently propose a two-component model to quantitatively fit the nonlinear HE,

$$\widetilde{R}(B,T) = R_0 f(B,T) + R_1 [1 - f(B,T)] \qquad (1)$$

where $\widetilde{R}(B,T)$ is the slope of the Hall resistivity at field $B$ and temperature $T$, $R_0$ is the initial Hall coefficient, $R_1$ is the Hall resistivity slope at high fields, and $f(B,T)$ is a weight function explicitly defined as magnetization dependent,

$$f(B,T) = \{1 + \exp[C(\overline{\mu} - \mu_C)]\}^{-1} \qquad (2)$$

where $\overline{\mu} = M/N \propto \chi H$ ($M$ is magnetization and $N$ is the number of Eu sites per unit volume) is the average magnetization per Eu site, and in the paramagnetic phase, is proportional to $B/(T-\theta)$. $\mu_C$ is the constant critical magnetic moment per Eu site at which the Hall resistivity slope switches and $C$ is a constant which defines the sharpness of the transition. The best fit to the data yields $\mu_C = 0.66\mu_B$ and $C = 6.8/\mu_B$, and the curve is shown in Fig. 2(c); the weight function is plotted in the inset. The results demonstrate that regardless of temperature, the transition in the Hall resistivity slope occurs when the average magnetic moment reaches only about 10% of the saturation moment ($7\mu_B$) for each $Eu^{2+}$.

To understand physically the existence of the constant $M_C$, we begin with the view of intrinsic electronic/magnetic inhomogeneities. Such non-chemical phase separation has been widely evidenced theoretically [2,14] and experimentally [15-17] in systems that display colossal magnetoresistance (CMR), or more generally, large negative magnetoresistance (negative MR). There are several possible reasons, including impurities, lattice distortions, and carrier-local moment exchange interactions, that may cause such inhomogeneities in a magnetic system. In particular, a picture of magnetic polarons successfully explains the observed giant negative MR in several rare earth chalcogenides [18]. Although this specific concept may not be readily applicable to $EuB_6$ since the system has negligible impurities which act as the localization centers [19], the electronic phase separation most likely originates from the exchange interactions between holes and local moments, which could result in large magnetic polarons [20], regular magnetic polarons [21], or even smaller entities only a few nanometers in diameter [2,22]. Regardless of the microscopic picture, at zero field the electronic inhomogeneities restrict the mobility of the strongly coupled holes, causing their localization. However, with an applied magnetic field and/or decreasing temperature, local moments begin to be aligned, increasing either the size of the phase-separated entities or their number. Eventually these more conducting regions coalesce, leading to *delocalization* of the holes. Such electronic inhomogeneities require strong coupling between charge carriers and local moments, thus the microscopic distribution of the more conducting entities is directly related to the overall magnetization, resulting in delocalization at a universal critical magnetization.

Our earlier results of spin polarization and transport measurements showed that the valence band undergoes a spontaneous band splitting during ferromagnetic ordering and concomitant delocalization of holes [12]. The observed Hall slope switching implies that such delocalization of the holes can also be induced in the paramagnetic phase at a critical magnetization attained by applying a sufficiently large magnetic field. The constants $R_0$ and $R_1$ in Eq. 1 then have unambiguous physical meanings: with a negligible AHE term, $R_0$ corresponds to an effective Hall coefficient in the hole-localized phase, and it predominantly reflects the density of conducting electrons; on the other hand, $R_1$ reflects the effective two-band carrier density of the electrons and holes in the hole-delocalized phase. This explains why $R_1$ is consistent with the Hall resistivity slope in the ferromagnetic phase, to which the two-band model has been successfully applied [12]. In addition, the weight function defined in Eq. 2 represents the probable fraction of the hole-localized regions and thus offers a quantitative measure of the electronic phase separation in the system.

With the carrier delocalization characterized by a $M_C$ far below the saturation value, the two consecutive transitions of EuB$_6$ [8-10] can be interpreted as follows: The higher temperature transition is a percolative transition from overlapping of the patches of a phase with higher electronic conductivity and magnetic ordering; at this temperature, the holes begin to be delocalized while the material is still paramagnetic. Global spontaneous alignment of local moments (ferromagnetic ordering) occurs at the lower transition temperature. The picture is consistent with earlier suggestions of charge delocalization [9, 23].

The charge delocalization can also be induced by an applied magnetic field at higher temperatures. Both of the zero-field transitions have signatures in the temperature dependences of resistivity (peaks in $d\rho/dT$) [9,10] and heat capacity [10]. Fig. 3(b) shows $\rho(T)$ at different magnetic fields for the EuB$_6$ crystal. A circle represents the transition temperature for each field calculated from the $M_C$ determined from the HE data, which is clearly consistent with the charge delocalization point identified from the corresponding resistivity curve (higher temperature peak of $d\rho/dT$). At zero field, the specific heat curve [8-10] shows a small kink at ~15 K and a much more pronounced feature at ~12 K. This is also consistent with our picture, since the carrier density is about $10^{-3}$ of the density of local moments, thus the entropy associated with the local moments is far larger than that for the electron gas. Moreover, the blue shift in the plasma frequency with decreasing temperature or with applied field derived from magneto-optical measurements [24] can also be explained in terms of charge delocalization, which leads to an effective increase in carrier number.

The weight function defined in Eq. 2 offers a quantitative measure of the dominance of the phase with carrier localization. Fig. 4(a) shows a phase diagram in the field-temperature plane evaluated from the weight function determined from the best fit shown in Fig. 2(c). The dashed line is calculated from $M_C$, which coincides with the best-fit line to the $B_C$ versus $T$ data. The solid line represents the ferromagnetic ordering of the local moments which is not sensitive to external field. In essence, the color measures the degree of the electronic phase separation and the dashed line represents the transition temperature/field for the percolative charge delocalization transition.

Our model for the nonlinear HE has only two necessary ingredients: strong exchange coupling between charge carriers and local magnetic moments and electronic phase separation at zero magnetic field. These two characteristics are quite common in transition metal and rare-earth metal compounds. Therefore, the magnetic field induced percolative carrier delocalization transition may be a common phenomenon in these systems reflected in their transport properties, including the HE. An intriguing example is the antiferromagnetic heavy Fermion metal YbRh$_2$Si$_2$ [7], in which a nonlinear Hall resistivity with a distinct slope change has been reported. The crossover field is linear with temperature and extrapolates to the Curie-Weiss temperature of the material [25]. Another notable example is the mixed-valence manganese perovskites, in which there is extensive theoretical [2,14] and experimental evidence [15-17] for intrinsic electronic inhomogeneities near the ferromagnetic ordering temperature. Ubiquitous switches in the Hall resistivity slope [26-30] unrelated to the conventional AHE have been observed. Various authors have noted that the nonlinear HE may be related to polaron conduction [28,29] or increasing carrier density [30], and excellent scaling of the Hall resistivities as functions of magnetization is evident [29]. These observations strongly imply that our picture and model for the nonlinear HE may be applicable to the manganites as well. As an example, we analyzed the data by Yang *et al.* on $Nd_{0.7}Sr_{0.3}MnO_3$ [27] based on the two-component model. With its relatively low $T_C$, systematic data of magnetization and HE were obtained [27] in the paramagnetic phase over a sufficiently broad temperature range. Excellent agreement with the model resulted from the analysis. Fig. 4(b) shows the switching fields of the Hall resistivity, which exhibit a linear temperature dependence that extrapolates to 225 K, consistent with the Curie temperature (222 K) of the material [27]. Furthermore, the best-fit of the data yields a constant critical magnetization of about 12% of the saturation magnetization of the material.

To summarize, the magnetic phase transitions in $EuB_6$ are reflected in dramatic behavior of its nonlinear HE: the changes in the Hall resistivity slope occur at a single constant critical magnetization in a wide range of temperatures. A two-component model characterized by this critical magnetization provides excellent fits and scaling to the HE data. Similar nonlinear HE has been observed in many other transition metal and rare-earth compounds and is well described by the model. We believe this type of nonlinear HE may be a common signature of magnetically-induced electronic phase separation in a wide variety of correlated electron materials and could offer a new quantitative probe for its study.

This work was supported in part by NSF grants DMR-0908625 (S.v.M. and P.X.) and DMR-0710492 and -0503360 (Z.F.). S.v.M. acknowledges several illuminating discussions with S. Wirth and E. Manousakis.

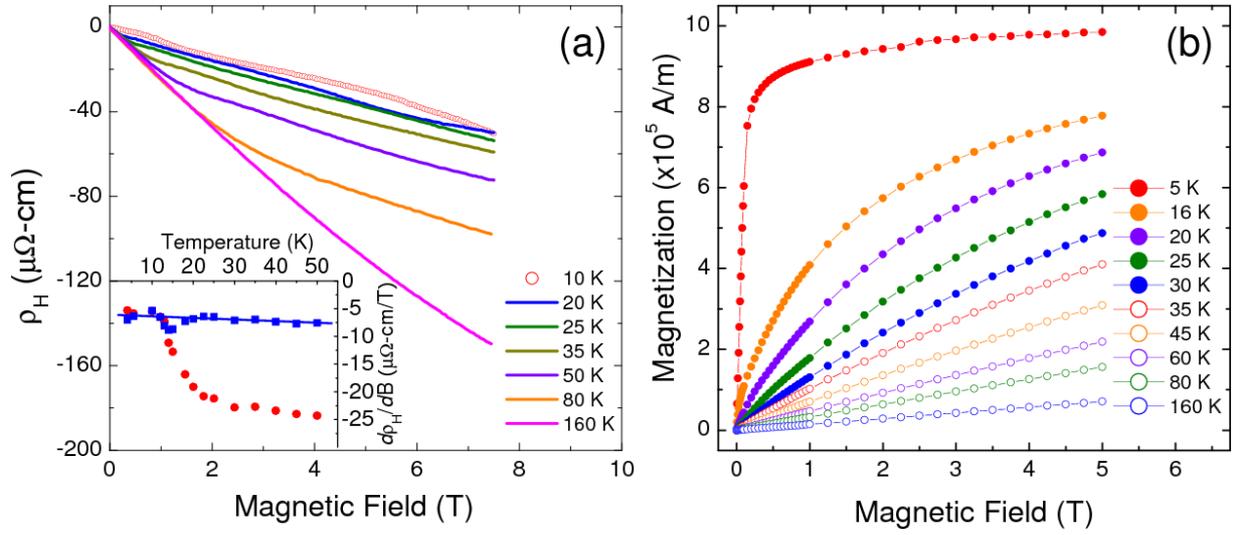

FIG. 1 (color online). (a) Hall resistivity of a EuB$_6$ crystal at selected temperatures. Solid lines are data for paramagnetic temperatures; the circles show the Hall resistivity for 10 K. The inset shows the Hall slopes at low field (initial Hall coefficients, solid circles) and high field (empty squares) for different temperatures. The line is a guide to the eye as described in the text. (b) Magnetization versus field for different temperatures.

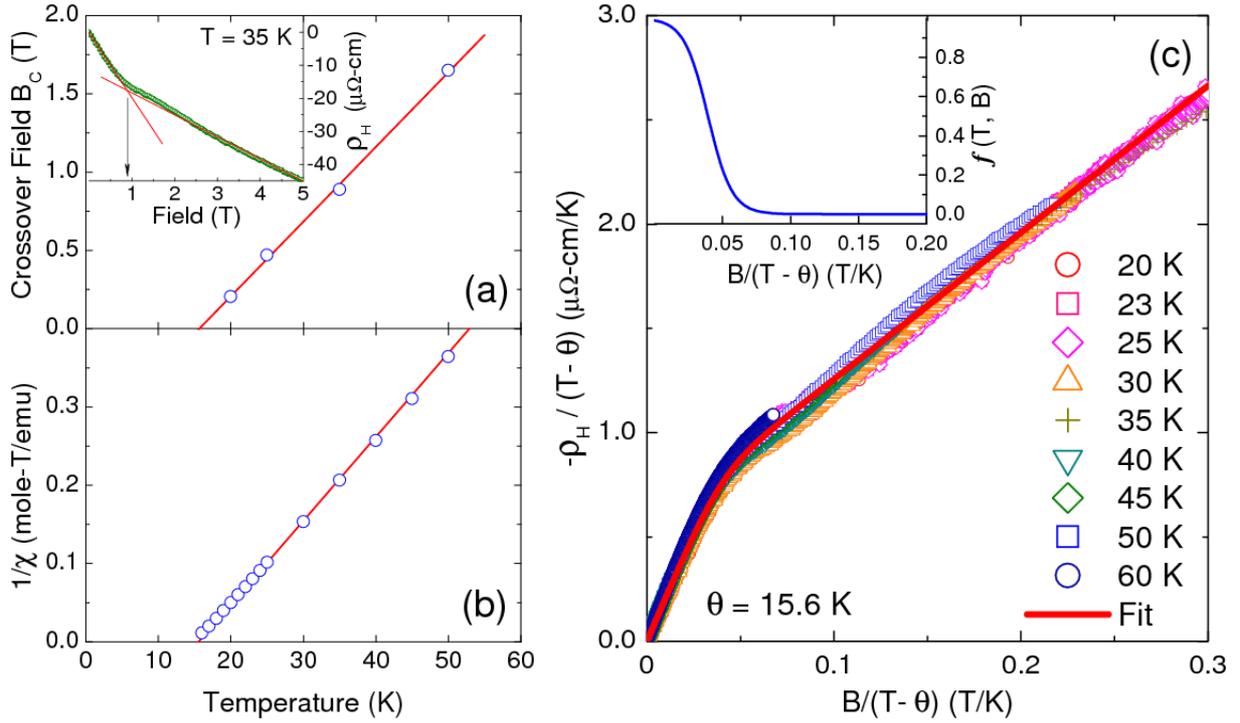

FIG. 2 (color online). (a) Temperature dependence of the switching field of the Hall resistivity. An example illustrating the determination of the switching field is shown in the inset. The solid line is a linear fit to the points. (b) Temperature dependence of the reciprocal magnetic susceptibility measured at 200 gauss. The Curie temperature is 15.53 K. (c) Symbols are rescaled Hall resistivities at different temperatures from 20 K to 60 K. The horizontal axis $B/(T-\theta)$ is proportional to magnetization in the paramagnetic phase. The solid curve is a fit to the data based on the model described in the text. The inset is the weight function used in the fit.

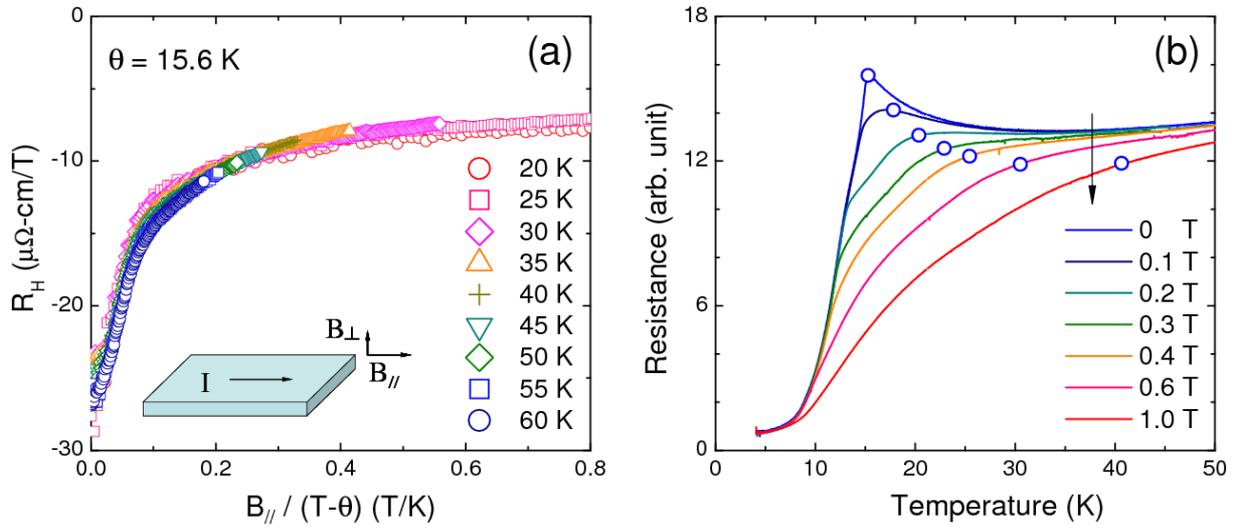

FIG. 3 (color online). (a) Low-field Hall coefficient as functions of magnetization induced by an in-plane magnetic field ($B_{//}$) at different temperatures. Small out-of-plane magnetic fields ($B_\perp$) were applied to obtain Hall signals. (b) Temperature dependence of resistivity at different magnetic fields. The circles label the transition temperature (see Fig. 2(a)) for each field calculated from the fit in Fig. 2(c).

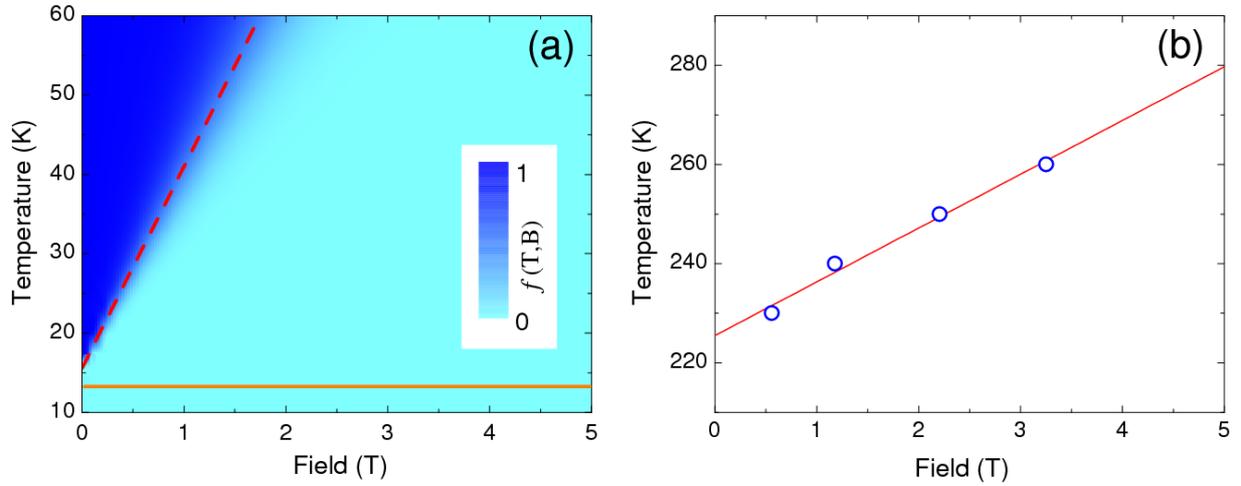

FIG. 4 (color online). (a) A phase diagram in the field-temperature plane evaluated from the weight function. The dashed line corresponds to carrier delocalization while the solid line corresponds to the ferromagnetic ordering of the local moments. (b) Crossover points determined from the nonlinear HE of $Nd_{0.7}Sr_{0.3}MnO_3$ (ref. 27). The line is a linear fit to the data which extrapolates to the Curie temperature of the material.

REFERENCES

1. S. Kivelson *et al.*, Rev. Mod. Phys. **75**, 1201 (2003).

2. E. Dagotto, T. Hotta, and A. Moreo, Phys. Rep. **344,** 1 (2001).

3. S. M. Watts *et al.*, Phys. Rev. B **61,** 9621 (2000).

4. C. L. Chien and C. R. Westgate, *The Hall Effect and Its Applications* (Plenum Press, New York, 1980).

5. T. F. Rosenbaum, S. B. Field, D. A. Nelson, and P. B. Littlewood, Phys. Rev. Lett. **54,** 241 (1985).

6. V. J. Goldman, M. Shayegan, and H. D. Drew, Phys. Rev. Lett. **57,** 1056 (1986).

7. S. Paschen *et al.*, Nature **432,** 881 (2004).

8. L. Degiorgi *et al.*, Phys. Rev. Lett. **79,** 5134 (1997).

9. S. Süllow *et al.*, Phys. Rev. B **62,** 11626 (2000).

10. R. R. Urbano *et al.*, Phys. Rev. B **70,** 140401(R) (2004).

11. R. G. Goodrich *et al.*, Phys. Rev. B **58,** 14896 (1998); M. C. Aronson *et al.*, Phys. Rev. B **59,** 4720 (1999).

12. X. H. Zhang, S. von Molnár, Z. Fisk, and P. Xiong, Phys. Rev. Lett. **100,** 167001 (2008).

13. Y. Shapira and T. B. Reed, Phys. Rev. B **5,** 4877 (1971).

14. L. P. Gor'kov and V. Z. Kresin, JETP Lett. **67,** 985 (1998).

15. R. D. Merithew *et al.*, Phys. Rev. Lett. **84,** 3442 (2000).

16. B. Raquet *et al.*, Phys. Rev. Lett. **84,** 4485 (2000).

17. S. von Molnár and J. M. D. Coey, Current Opinions in Solid State & Material Science **3,** 171 (1998).

18. S. von Molnár and S. Methfessel, J. Appl. Phys. **38,** 959 (1967).

19. T. Kasuya and A. Yanase, Rev. Mod. Phys. **40,** 684 (1968).

20. R. N. Bhatt, M. Berciu, M. P. Kennett, and X. Wan, J. Supercond. **15,** 71 (2002); A. Kaminski and S. Das Sarma, Phys. Rev. Lett. **88,** 247202 (2002).

21. U. Yu and B. I. Min, Phys. Rev. Lett. **94**, 117202 (2005).

22. S. Bobler *et al.*, Europhys. Lett. **83**, 17009 (2008).

23. M. L. Brooks *et al.*, Phys. Rev. B **70**, 020401(R) (2004).

24. S. Broderick *et al.*, Phys. Rev. B **65,** 121102(R) (2002).

25. P. Gegenwart *et al.*, Phys. Rev. Lett. **89,** 056402 (2002).

26. M. B. Salamon and M. Jaime, Rev. Mod. Phys. **73,** 583 (2001).

27. H. C. Yang, L. M. Wang, and H. E. Horng, Phys. Rev. B **64,** 174415 (2001).

28. S. H. Chun, M. B. Salamon, and P. D. Han, Phys. Rev. B **59,** 11155 (1999).

29. Y. Lyanda-Geller *et al.*, Phys. Rev. B **63,** 184426 (2001).

30. H. Imai, Y. Shimakawa, Y. V. Sushko, and Y. Kubo, Phys. Rev. B **62,** 12190 (2000).